\documentclass[%
 reprint,
 amsmath,amssymb,
 longbibliography,
 aps,
prl
]{revtex4-1}

\usepackage[normalem]{ulem}
\usepackage{graphicx}
\usepackage{bm}
\usepackage{xcolor}
\usepackage{hyperref}
\usepackage{amsmath}
\usepackage[utf8]{inputenc}
\usepackage{ulem}
\usepackage{xr}

\newcommand{\mean}[1]{\mathrm{Mean}\left(#1\right)}
\newcommand{\var}[1]{\mathrm{Var}\left(#1\right)}

\newcommand{\varasy}[1]{\mathrm{Var}^{\textrm{\tiny{asy}}}\!\left(#1\right)}

\newcommand{\meannum}[1]{\mathrm{Mean}^{\textrm{\tiny{num}}}\!\left(#1\right)}
\newcommand{\varnum}[1]{\mathrm{Var}^{\textrm{\tiny{num}}}\!\left(#1\right)}

\def\maxnt{\mathrm{Max}^{N}_t}
\def\envnt{\mathrm{Env}^{N}_t}
\def\snt{\mathrm{Sam}^{N}_t}

\makeatletter
\newcommand{\hathat}[1]{%
\begingroup%
  \let\macc@kerna\z@%
  \let\macc@kernb\z@%
  \let\macc@nucleus\@empty%
  \hat{\raisebox{.2ex}{\vphantom{\ensuremath{#1}}}\smash{\hat{#1}}}%
\endgroup%
}
\makeatother

\begin{document}

\title{Anomalous Fluctuations of Extremes in Many-Particle Diffusion}
\author{Jacob B. Hass$^*$, Aileen N. Carroll-Godfrey$^*$, Ivan Corwin$^\dagger$, Eric I. Corwin$^*$}
\affiliation{$^*$Department of Physics and Materials Science Institute, University of Oregon, Eugene, Oregon 97403, USA. \\ $^\dagger$Department of Mathematics, Columbia University, New York, New York 10027, USA.}
\date{\today}

\begin{abstract}
In many-particle diffusions, particles that move the furthest and fastest can play an outsized role in physical phenomena. A theoretical understanding of the behavior of such extreme particles is nascent. A classical model, in the spirit of Einstein's  treatment of single-particle diffusion, has each particle taking independent homogeneous random walks. This, however, neglects the fact that all particles diffuse in a common and often inhomogeneous environment that can affect their motion. A more sophisticated model treats this common environment as a space-time random biasing field which influences each particle's independent motion. While the bulk (or typical particle) behavior of these two models has been found to match to high degree, recent theoretical work of Barraquand, Corwin and Le Doussal on a one-dimensional exactly solvable version of this random environment model suggests that the extreme behavior is quite different between the two models. We transform these asymptotic (in system size and time) results into physically applicable predictions. Using high precision numerical simulations we reconcile different asymptotic phases in a manner that matches numerics down to realistic system sizes, amenable to experimental confirmation. We characterize the behavior of extreme diffusion in the random environment model by the presence of a new phase with anomalous fluctuations related to the Kardar-Parisi-Zhang universality class and equation.

\end{abstract}

\maketitle

\noindent\emph{Introduction---}
Our world is fueled by outliers. Information in signals is carried by the leading edge \cite{saxtonSingleparticleTrackingApplications1997a,pintoPhysicsTypeIa2000, hoflingAnomalousTransportCrowded2013a, ghoshNonuniversalTracerDiffusion2015, manzoReviewProgressSingle2015, iyer-biswasFirstPassageProcesses2016,metzlerNonBrownianDiffusionLipid2016c,  shenSingleParticleTracking2017}.
A viral or bacterial infection is spread by the first few pathogens to enter a host and the first host to enter a new region \cite{hufnagelForecastControlEpidemics2004a,kaoVirusDiffusionIsolation2006b}.
A species is evolved by the fittest mutations \cite{ewensMathematicalPopulationGenetics2012, metzlerFirstpassagePhenomenaTheir2014,waxmanDiffusionEquationRandom2017}. Scientific revolution is sparked by the first new idea. In all of these contexts the precipitating action is driven by the extremes among a great number of agents (varying from $N\sim 10^2$ to $N\sim 10^{60}$ depending on the context) evolving in a complex but shared environment. How does the nature of the shared environment affect these outlier behaviors? Conversely, can we infer the nature of the shared environment from the behavior of these outliers?  Despite their obvious importance, these overarching questions are still unanswered.

The classical model for many-particle diffusion as independent homogeneous random walks provides an easily calculable solution, but entirely neglects the effects of the shared and likely inhomogeneous environment. This model is the basis for diffusion coefficients~\cite{einsteinUberMolekularkinetischenTheorie1905b,einsteinZurTheorieBrownschen1906a, einsteinTheoretischeBemerkungenUber1907a}, which succinctly describe the behavior of typical particles in a many-particle diffusion. A more sophisticated model treats the shared environment as a space-time random biasing field with short-range space-time correlations. Each particle thus articulates independent random walks subject to forcing by the common biasing field. While this refined model does not affect typical particle diffusion behavior \cite{rassoul-aghaQuenchedFreeEnergy2013}, it drastically impacts the behavior of extreme particles. In this work, we provide predictions for the behavior of extreme particles moving in a random and inhomogenous environment.
We find that the variance in the position of the extreme particle is a robust and sensitive measurement of the nature of the environment and show how this variance can be understood as the sum of two contributions: the randomness present in the environment, and the sampling of random walks in that environment.  We show that by subtracting out the variance due to sampling we can produce direct measurements of the environment, inaccessible from measurements of the motion of a typical particle or of the bulk. This residual environmental variance is characterized by a novel power law that we demonstrate holds even when the number of particles is as small as a few hundred.

\begin{figure}[h]
	\includegraphics[width=.9\columnwidth]{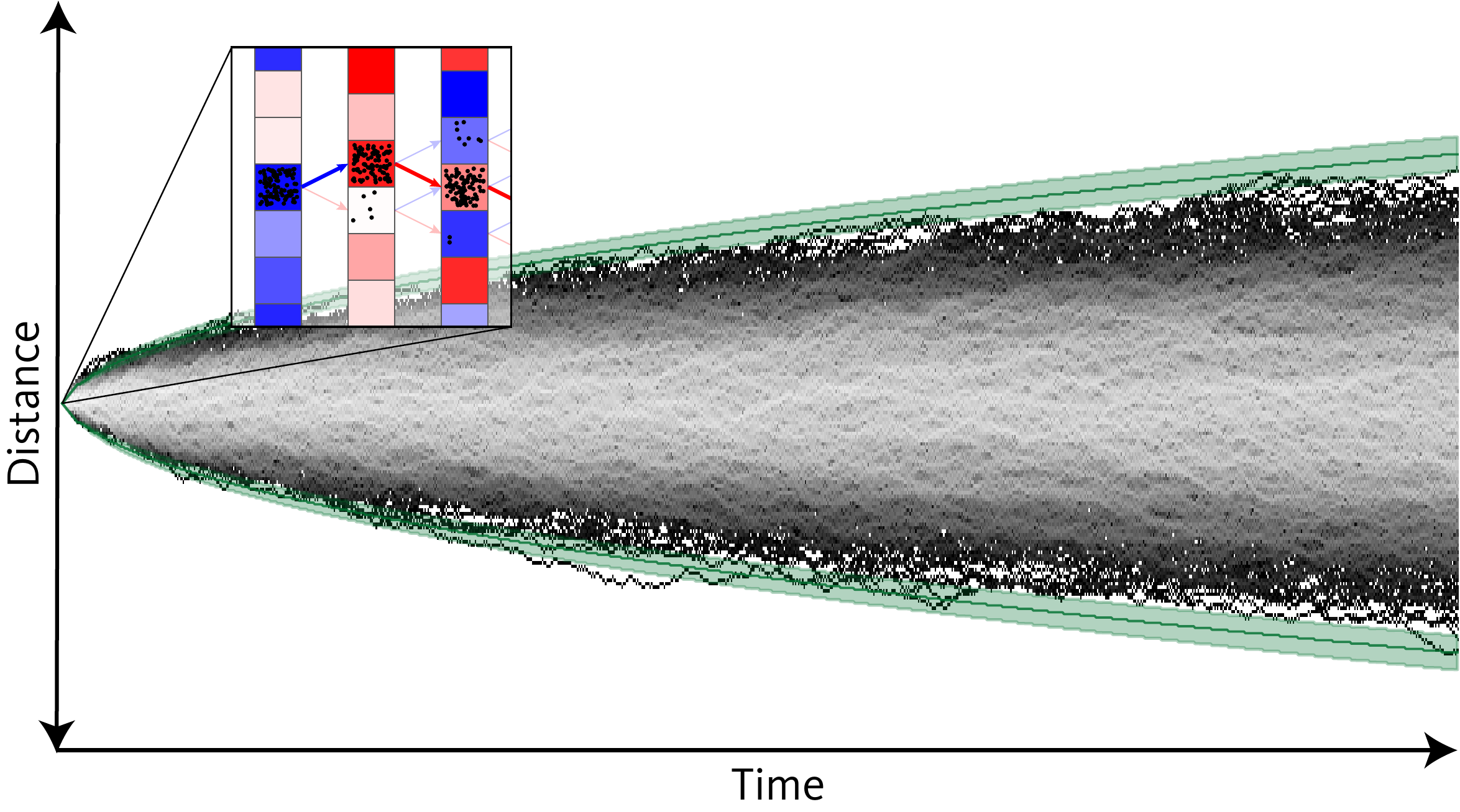}
  \caption{A system of $N=10^5$ particles evolving in a given random environment. The heat map records the site occupancy density. We also plot in green the asymptotic theory mean location for the maximum particle location. Around this is a shaded region with a width of two standard deviations based on the asymptotic theory variance. This region generally contains the extreme-most particle over time. The zoomed-in inset shows the spatial locations of $N=10^2$ particles over time. Color indicates the bias (red is biased down and blue is biased up) and is chosen independently at each space-time box. The location of particles within each box is chosen for ease of visualization.}
  \label{fig:BCModel}
\end{figure}

\medskip\noindent\emph{Background---}
Building on observations by Brown~\cite{brownXXVIIBriefAccount1828a, brownXXIVAdditionalRemarks1829a} from 1827,  Einstein~\cite{einsteinUberMolekularkinetischenTheorie1905b,einsteinZurTheorieBrownschen1906a, einsteinTheoretischeBemerkungenUber1907a} (along with Langevin~\cite{langevinTheorieMouvementBrownien1908a}, Sutherland~\cite{sutherlandLIIViscosityGases1893,sutherlandLXXVDynamicalTheory1905} and Smoluchowski~\cite{vonsmoluchowskiZurKinetischenTheorie1906, vonsmoluchowskiNotizUiberBerechnung1915}) proposed a theory of diffusion based on modeling particles by independent random walks with variance controlled by a diffusion coefficient intrinsic to the particle/environment pair. Soon after, Perrin experimentally verified Einstein's statistical predictions \cite{perrinMouvementBrownienRealite1909a, perrinMouvementBrownienMolecules1910a}.

Probing the effectiveness and limitations of Einstein's diffusion model has remained a challenge. On short time scales, particle motion is ballistic, dominated by inertia~\cite{uhlenbeckTheoryBrownianMotion1930a,huangDirectObservationFull2011,hammondDirectMeasurementBallistic2017,lukicDirectObservationNondiffusive2005, franoschResonancesArisingHydrodynamic2011, kheifetsObservationBrownianMotion2014}.
Many physically relevant situations require the addition of new concepts to accurately model them.
Certain diffusive processes are better modeled by Levy flights \cite{wangWhenBrownianDiffusion2012a} or other types of anomalous diffusions~\cite{bouchaudAnomalousDiffusionDisordered1990a, metzlerBrownianMotionFirstpassage2019a} instead of simple random walks. Other work has focused on active particles which inject energy into their environment~\cite{ramaswamyMechanicsStatisticsActive2010b,kanazawaLoopyLevyFlights2020a}. Further, in environments which are slowly mixing, Einstein's theory may also break down due to the presence of quenched disorder \cite{zangiFrequencydependentStokesEinsteinRelation2007, wangWhenBrownianDiffusion2012a}. Unlike the above deviations from the classical model, our approach is intended to describe generic many-particle diffusions.

The random walk in random environment (RWRE) model goes back to~\cite{chernovReplicationMulticomponentChain1967, temkinOnedimensionalRandomWalks1972} (see also~\cite{havlinDiffusionDisorderedMedia1987, bolthausenTenLecturesRandom2002,sznitmanTopicsRandomWalks2004, oferRandomWalksRandom2004, hughesRandomWalksRandom1995}) and comes in two types -- long-range~\cite{kestenLimitLawRandom1975,sinaiLimitingBehaviorOneDimensional1983a,bouchaudClassicalDiffusionParticle1990a, bouchaudAnomalousDiffusionDisordered1990a, burlatskyTransientRelaxationCharged1998, ledoussalRandomWalkersOnedimensional1999} and short-range~\cite{richardsonAtmosphericDiffusionShown1926, hentschelRelativeDiffusionTurbulent1984, bouchaudDiffusionLocalizationWaves1990,  chertkovAnomalousScalingExponents1996, bernardAnomalousScalingNPoint1996, jullienRichardsonPairDispersion1999a, balkovskyIntermittentDistributionInertial2001} temporally correlated environments. We focus here on the latter. In this context, typical RWRE particles behave like Brownian motion, matching the behavior from Einstein's model~\cite{rassoul-aghaAlmostSureInvariance2005, deuschelQuenchedInvariancePrinciple2016}.
The motion of atypical particles is controlled by large deviations of the RWRE's transition probability as first studied in ~\cite{balazsRandomAverageProcess2006}.

Barraquand and Corwin~\cite{barraquandRandomwalkBetadistributedRandom2017a} discovered the exactly solvable Beta RWRE discussed extensively below and uncovered a remarkable connection between its large deviations for times of order $\log(N)$ and the statistics of the Kardar-Parisi-Zhang (KPZ) universality class~\cite{corwinKardarParisiZhang2012a, quastelOneDimensionalKPZEquation2015a}, namely the Gaussian Unitary Ensemble (GUE) Tracy-Widom distribution~\cite{tracyLevelspacingDistributionsAiry1993}. Soon after~\cite{ledoussalDiffusionTimedependentRandom2017} recognized that a phase transition should occur in the $(\log(N))^2$ time frame while~\cite{barraquandModerateDeviationsDiffusion2020a} discovered that in this frame the GUE Tracy-Widom distribution is replaced by the KPZ equation one-point distribution~\cite{kardarDynamicScalingGrowing1986a,sasamotoOneDimensionalKardarParisiZhangEquation2010,calabreseFreeenergyDistributionDirected2010,dotsenkoBetheAnsatzDerivation2010,amirProbabilityDistributionFree2011}. See~\cite{sabotRandomWalksDirichlet2017,balazsLargeDeviationsWandering2019, barraquandLargeDeviationsSticky2020, barraquandLargeDeviationsSticky2020, brockingtonBetheAnsatzSticky2021, oviedoSecondOrderFluctuations2021,korotkikhHiddenDiagonalIntegrability2022,krajenbrinkCrossoverMacroscopicFluctuation2022} for further developments. The recursion relation \eqref{eq:kolmogorov} for RWRE transition probabilities solves a discrete version of the multiplicative noise stochastic heat equation (mSHE)
\begin{equation}
\partial_t Z(x,t) = \frac{1}{2} \partial_x^2 Z(x,t) + \xi(x,t)Z(x,t)
\end{equation}
with $\xi$ space-time white noise. The logarithm of the mSHE $h(x,t)=\log Z(x,t)$ solves the KPZ equation
\begin{equation}\label{eq:KPZ}
\partial_t h(x,t) = \frac{1}{2} \partial_x^2 h(x,t) + \frac{1}{2}(\partial_x h(x,t))^2 + \xi(x,t).
\end{equation}
Hence, large deviations for RWREs, in particular beyond the solvable model and even in experimental settings, may relate to the KPZ equation and its universality class -- especially in light of the rich canon of work on KPZ universality in various contexts using theoretical~\cite{albertsIntermediateDisorderRegime2010, corwinKardarParisiZhang2012a}, numerical~\cite{PhysRevA.45.638, prolhacHeightDistributionKPZ2011}, and experimental~\cite{halpin-healyKPZCocktailShakenNot2015a} methods. The KPZ connection is  quite useful since its statistics and power-laws are well studied.

\medskip
\noindent\emph{Models for diffusion---}
Although physical diffusion is continuous in time and (typically) occurs in three-dimensional space, here we work with discrete models in one spatial dimension. The principal reason for this choice is that it is the setting for the exactly solvable  Beta RWRE~\cite{barraquandRandomwalkBetadistributedRandom2017a} (a continuous {\it sticky Brownian motion} limit of this model exists~\cite{barraquandLargeDeviationsSticky2020}) that will enable us to compare numerical results to exact theoretical predictions.
Beyond that, discretization is common for numerical simulations and higher dimensions are more challenging numerically due to anisotropy issues arising from the choice of lattice and due to the lack of exactly solvable models, c.f. \cite{ledoussalDiffusionTimedependentRandom2017}. In real diffusion in a common environment, there will be length and time scales on which the environment decorrelates.  Our discrete model can be thought of as coarse-graining the environment in space and time onto a lattice and thus we do not expect discrete and continuous models to differ greatly for long-times and large-scales. Our model ignores any higher order interactions as we expect them to be less present in the behavior of extreme particles, for which the local density is necessarily low. Additionally, there are physical settings where particles take discrete states~\cite{fellerDiffusionProcessesGenetics1951, moranRandomProcessesGenetics1958} or evolve in quasi-one-dimensional spaces~\cite{pollardGaseousSelfDiffusionLong1948,ahmadiDiffusionQuasionedimensionalChannels2017}.

We study the {\it Beta RWRE} introduced in~\cite{barraquandRandomwalkBetadistributedRandom2017a} (see Fig.~\ref{fig:BCModel}). We model the environment by a collection $\mathbf{B}= \big\{B(x,t):x\in \mathbb{Z},t\in \mathbb{Z}_{\geq 0}\big\}$ of independent identically distributed random variables all drawn from the uniform distribution on $[0,1]$. At time $t=0$ we start with $N$ particles all at site $0$. Given an instance of the environment $\mathbf{B}$ the particles proceed as follows. Each particle at $x$ and $t$ independently flips the same weighted coin which has probability $B(x,t)$ of heads (moving the particle to site $x+1$ at time $t+1$) and $1-B(x,t)$ of tails (moving to $x-1$ instead). Thus, while particles do not interact with each other, those at the same place and time are all influenced by the common environment.

This model is exactly solvable when $B(x,t)$ are distributed according to the Beta distribution, $Beta(\alpha, \beta)$~\cite{barraquandRandomwalkBetadistributedRandom2017a}. For simplicity, we focus on the special case $\alpha=\beta=1$ corresponding to the uniform distribution. The classical simple symmetric random walk (SSRW) model arises in the limit $\alpha=\beta\to \infty$ where all $B(x,t)\equiv 1/2$ and the environment is deterministic.

We focus on the behavior of the right-most particle at time $t$. We denote this by $\maxnt$, with $N$ the number of particles in the system. Two types of randomness affect $\maxnt$: that of the environment and that of sampling the random walks in that environment. The effect of the environment is via the transition probability $p_{\mathbf{B}}(x,t)$, the probability that a single random walker initially at $0$ will end up at $x$ at time $t$ for a given environment $\mathbf{B}$. This satisfies the recursion relationship, 
\begin{align} \label{eq:kolmogorov}
 \begin{split}
  p_{\mathbf{B}}(x,t) = & p_{\mathbf{B}}(x-1,t-1)B(x-1,t-1) +\\
  & p_{\mathbf{B}}(x+1,t-1)\big(1-B(x+1,t-1)\big)
 \end{split}
\end{align}
with initial condition $p_{\mathbf{B}}(0,0) = 1$ and $p_{\mathbf{B}}(x \neq 0,0) = 0$. Since each random walker is independent, conditional on the environment, the distribution of the ensemble of $N$ walks is determined by $p_{\mathbf{B}}(x,t)$. Given the environment $\mathbf{B}$, the probability that a single random walker is at or above $x$ at time $t$ is given by the tail probability, $P_{\mathbf{B}}(x,t) = \sum_{y\geq x} p_{\mathbf{B}}(y,t)$. This and the independence of random walkers, conditional on the environment, imply
\begin{equation}\label{eq:maxpb}
\mathrm{Prob}_{\mathbf{B}}(\maxnt \leq x) = \big(1-P_{\mathbf{B}}(x,t)\big)^N,
\end{equation}
where the left-hand side is the probability, given the environment $\mathbf{B}$, that $\maxnt \leq x$.

We study how $\maxnt$ varies upon sampling a new environment and random walkers therein. Eq. \eqref{eq:maxpb} suggests that a good proxy for $\maxnt$ is the location $\envnt$ of the $1/N$-quantile of $P_{\mathbf{B}}(x,t)$, i.e., $\envnt$ equals the maximal $x$ such that $P_{\mathbf{B}}(x,t)>1/N$. Notice that $\envnt$ only accounts for the variation due to the environment. The variation due to sampling in that environment is denoted $\snt$ and defined by $\maxnt = \envnt + \snt$. We use the notation $\mean{\bullet}$ and $\var{\bullet}$ for the mean and variance of a quantity $\bullet$ (e.g. $\maxnt, \envnt,\snt$) averaged over both the environment and the sampling of random walkers in that environment.

\begin{figure}[h]
 \includegraphics[width=\columnwidth]{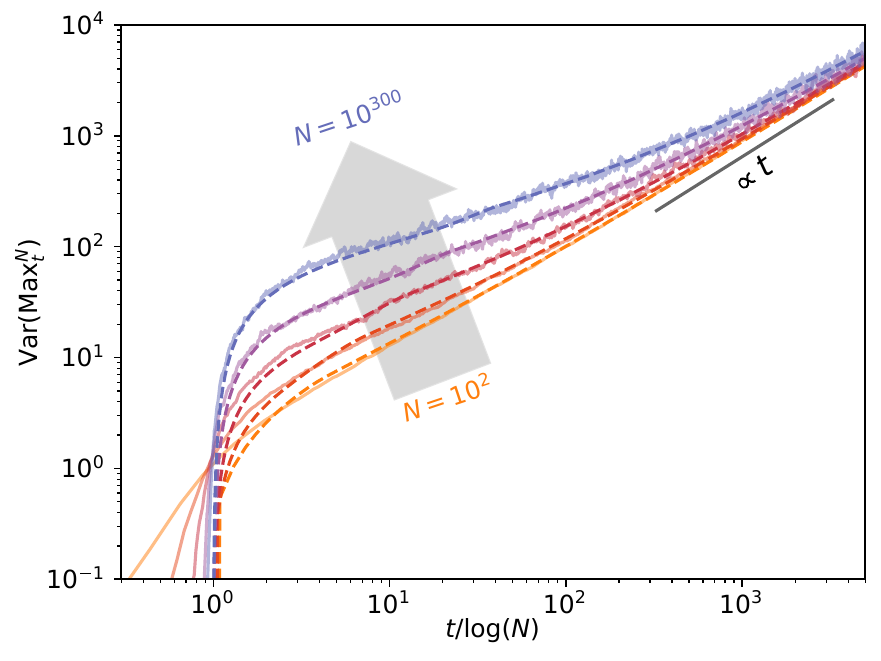}
 \caption{Plots of $\varnum{\maxnt}$ (solid) computed over 10000, 5000, 1000, 500 and 500 environments (respectively) and  $\varasy{\maxnt}$ (dashed) for $N=10^2, 10^{7}, 10^{24}, 10^{85}, 10^{300}$}
 \label{fig:MaxVar}
\end{figure}

\begin{figure}[h]
	\includegraphics[width=\columnwidth]{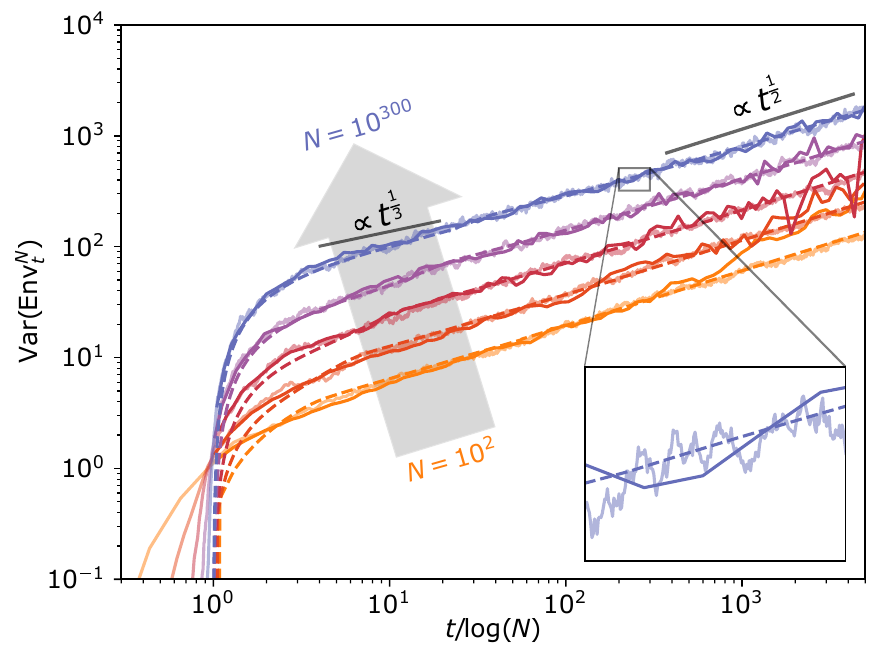}
	\caption{Plots of $\varnum{\envnt}$ (transparent solid) computed over 500 environments, $\varasy{\envnt}$ (dashed), and $\varnum{\maxnt} - \varasy{\snt}$ (dark solid) smoothed in each $1/25^{th}$ of a decade for $N=10^2, 10^{7}, 10^{24}, 10^{85},10^{300}$. The three curves agree as shown in the zoomed-in inset.}
	\label{fig:QuantileVar}
\end{figure}

\medskip\noindent\emph{Numerical Methods---}
We numerically simulate our models for system sizes varying from $N=10^2$ to $N=10^{300}$. We consider such large and physically unrealistic system sizes like $10^{300}$ in order to see how asymptotic theory applies for as wide a  range as possible of finite system sizes. We evolve the system for times from $t=0$ to $t= 5000 \log(N)$.  As explained below, $\log(N)$ and $(\log(N))^2$ set key timescales and our range of times ensure that for all choices of $N$, we encompass these scales. We simulate such large systems by tracking occupation variables instead of individual particle trajectories. In particular, if there are $N(x,t)$ particles at site $x$ at time $t$, then the number that move to site $x+1$ is binomially distributed with $N(x,t)$ samples and success probability $B(x,t)$ (the remainder move to site $x-1$). We sample these binomial distributions utilizing quadruple-precision floating point numbers and making approximations to the binomial distribution when dealing with sizes beyond our precision limits, as described in \cite{SeeSupplementalMaterial}. The rightmost particle location (identified by the maximal $x$ with $N(x,t)\geq 1$) at each time represents a sample of $\maxnt$. By repeatedly sampling new environments along with random walk occupation variables $N(x,t)$ therein we numerically measure $\var{\maxnt}$. To distinguish from the true value we denote this numerically measured variance by $\varnum{\maxnt}$ and plot it in Fig. \ref{fig:MaxVar}. In like fashion, we measure $\varnum{\envnt}$ for each sampled environment by using Eq. \ref{eq:kolmogorov} to compute $p_{\mathbf{B}}(x,t)$.  Fig. \ref{fig:QuantileVar} shows $\varnum{\envnt}$ as a function of time (see \cite{SeeSupplementalMaterial} for $\meannum{\maxnt}$ and $\meannum{\envnt}$). The data presented in Fig. \ref{fig:MaxVar} and \ref{fig:QuantileVar} took approximately three weeks to run in parallel on 500 cores of the University of Oregon high performance computing cluster, Talapas.

\medskip\noindent\emph{Asymptotic Theory Results---}
We describe asymptotic results on the behavior of $\maxnt$, $\envnt$ and $\snt$ as both $N$ and $t$ tend to infinity in different limits. Given a fixed relationship between $t$ and $\log(N)$ such as $t/\log(N)=\hat{t}$ or  $t/\log(N)^2=\hathat{t}$ for $\hat{t}$ or $\hathat{t}$ fixed, we write $f(N,t)\gg g(N,t)$ if $f(N,t)/g(N,t)$ tends to infinity as $N$ and $t$ do subject to their relationship. We use the notation $\varasy{\bullet}$ to denote the asymptotic theory formula for the variance of $\bullet$, interpolated back to finite $N$ and $t$. SSRW theory follows from Stirling's formula while asymptotic results for the RWRE rely on tools from quantum integrable systems~\cite{barraquandRandomwalkBetadistributedRandom2017a, barraquandModerateDeviationsDiffusion2020a,krajenbrinkCrossoverMacroscopicFluctuation2022} and are derived first for $\envnt$ and then for $\maxnt$ and $\snt$.

\noindent\textbf{SSRW $\boldsymbol{\maxnt}$:}
For $t/\log(N)=\hat{t}$ with fixed $\hat{t} < (\log 2)^{-1}$, we have $N\gg 2^t$ and hence with very high probability every reachable site in the lattice at time $t$ is occupied, hence $\var{\maxnt}\approx 0$. When $\hat{t} > (\log 2)^{-1}$, we show  in \cite{SeeSupplementalMaterial} that $\maxnt$ is asymptotically a Gumbel random variable. For $\hat{t}$ large, $\varasy{\maxnt} \approx \frac{\pi^2}{12} \frac{t}{\log(N)}$.


\noindent\textbf{RWRE $\boldsymbol{\envnt}$:}
For $t/\log(N)=\hat{t}$ with fixed $\hat{t}<1$, $\var{\envnt}\approx 0$. To see this, note that $P_\mathbf{B}(t,t) = B_{0,0} \cdots B_{t-1,t-1}$. Taking logs and applying the central limit theorem shows that $\log \big(P_\mathbf{B}(t,t)\big) \approx -t  + t^{1/2} G$ for $G$ a standard Gaussian. This implies that $P_\mathbf{B}(t,t) \approx e^{-t}\gg 1/N$. Thus the RWRE stops saturating the lattice when $t= \log(N)$ plus a order $(\log(N))^{1/2}$ Gaussian fluctuation. For the SSRW this happens at time $\log_2(N)$ plus order one fluctuations.

$\var{\envnt}$ displays two asymptotic regimes. For fixed $t/\log(N)=\hat{t}>1$, $\var{\envnt}$ takes  asymptotic form,
\begin{equation}\label{eqlogN}
V_1(N, t):=\Big(\frac{\log(N)}{t}\Big)^{2/3} \sigma_{\chi}^2 \frac{2^{2/3}\big(1-\frac{\log(N)}{t}\big)^{4/3}}{1- \big(1- \frac{\log (N)}{t}\big)^2},
\end{equation}
where $\sigma_{\chi}^2\approx~0.813$ is the variance of the GUE Tracy-Widom distribution \cite{prahoferUniversalDistributionsGrowth2000, tracyLevelspacingDistributionsAiry1993}. As shown in \cite{SeeSupplementalMaterial}, this follows from the result of \cite{barraquandRandomwalkBetadistributedRandom2017a}: For $v\in(0,1)$, $\log P_\mathbf{B}(vt,t)=-t I(v) + t^{1/3} \sigma(v)\chi_t$ where $I(v) = 1-\sqrt{1-v^2}$, $\sigma(v) = (2I(v)^2/(1-I(v)))^{1/3}$ and $\chi_t$ is random converging to the GUE Tracy-Widom distribution as $t$ goes to infinity.

For $t/(\log(N))^2\!=\!\hathat{t}$,  $\var{\envnt}$ takes asymptotic form
\begin{equation}\label{eqlogNsq}
V_2(N, t):=\frac{t}{2\log(N)} \cdot \var{h\Big(0,\frac{4 (\log(N))^2}{t}\Big)},
\end{equation}
where $h(0,s)$ is the height at $0$ and time $s$ of the {\it narrow wedge} solution to the KPZ equation \eqref{eq:KPZ}. As shown in \cite{SeeSupplementalMaterial}, this follows from \cite{barraquandModerateDeviationsDiffusion2020a}: For $v\in (0,\infty)$,
$
\log P_\mathbf{B}(vt^{3/4},t) \approx -\frac{v^2t^{1/2}}{2} -\log(t)/4+\log(v) - v^4/12 + h(0,v^4).
$

Interpolating between these regimes, and extrapolating past $(\log(N))^2$ (see also \cite{krajenbrinkCrossoverMacroscopicFluctuation2022}) we find two power-laws,
\begin{equation}\label{eq:varqnt}
\varasy{\envnt} \approx
    \begin{cases}
    \sigma_{\chi}^2 (\frac{\log(N)}{2})^{\frac{1}{3}} t^{\frac{1}{3}}& 1 \ll \frac{t}{\log(N)}\ll \log(N),\\
    \frac{1}{2}\pi^{\frac{1}{2}} t^{\frac{1}{2}}& \frac{t}{\log(N)}\gg \log(N).
    \end{cases}
\end{equation}
For finite $N$ and $t$ these regimes have a gentle crossover that we capture by setting
$
\varasy{\envnt} := I(N,t) V_1(N, t)+\big(1-I(N,t)\big) V_2(N, t)
$
where $I(N,t) := \frac{1}{2} \cdot \left(1-\text{erf}\left(\frac{t-(\log(N))^{3/2}}{(\log(N))^{4/3}}\right)\right)$ (with  $\text{erf}(x)= 2/\sqrt{\pi} \int_{0}^{x} e^{-t^2}dt$ the error function) interpolates from $1$ to $0$ over an interval of width $(\log(N))^{4/3}$ around $(\log(N))^{3/2}$.

\noindent\textbf{RWRE $\boldsymbol{\snt}$ and $\boldsymbol{\maxnt}$:}
We identify the additional contribution from sampling the many-particle diffusion given an environment.
Using Eq. \eqref{eq:maxpb} and Taylor expansion of the results of \cite{barraquandRandomwalkBetadistributedRandom2017a} and \cite{barraquandModerateDeviationsDiffusion2020a} quoted above, \cite{SeeSupplementalMaterial} shows that for $t/\log(N)=\hat{t}>1$ the sample fluctuation $\snt$ is of Gumbel type with variance
\begin{equation}\label{eq:varnst}
\varasy{\snt} = \frac{\pi^2}{6} \frac{\big(\frac{t}{\log(N)} -1\big)^2}{2\frac{t}{\log(N)} -1}\approx \frac{\pi^2}{12} \frac{t}{\log(N)}
\end{equation}
as $\hat{t}$ grows. This limit matches the behavior of the SSRW model. In \cite{SeeSupplementalMaterial} we also show that $\snt$ is asymptotically independent of $\envnt$, thus
\begin{equation}\label{eq:varadd}
\var{\maxnt} \approx \var{\envnt} + \var{\snt}.
\end{equation}

\begin{figure}[h]
  \includegraphics[width=\columnwidth]{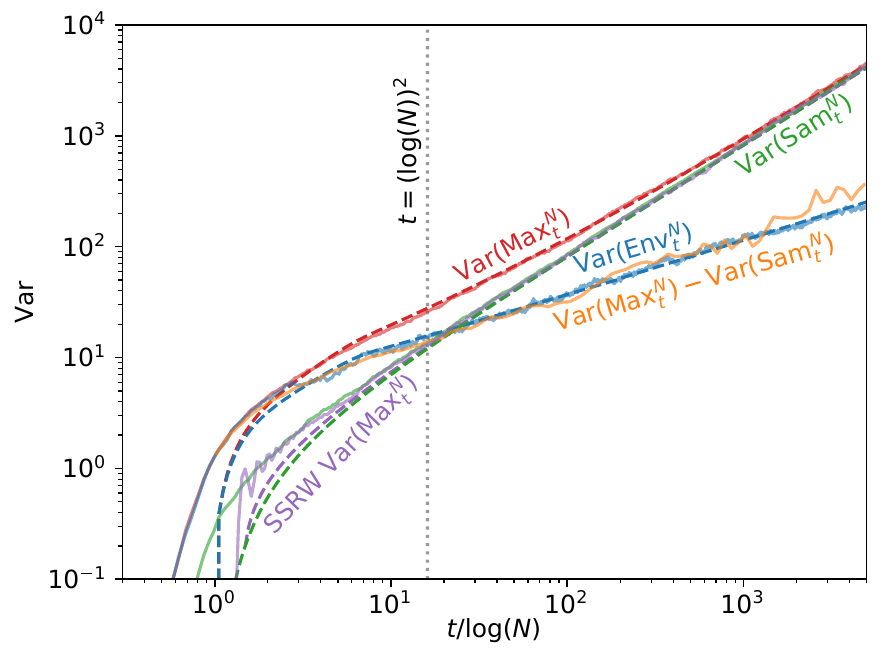}
  \caption{Variance of the maximal particle $\var{\maxnt}$ (red), environment $\var{\envnt}$ (blue), and sampling $\var{\snt}$ (green) for RWRE, and variance of the maximal particle  $\var{\maxnt}$ (purple) for SSRW, all for $N=10^7$. Dashed lines are $\varasy{\bullet}$, while solid lines are $\varnum{\bullet}$. $\varnum{\maxnt} - \varasy{\snt}$ (orange; smoothed as in Fig. \ref{fig:QuantileVar}) closely matches the environment curve (blue).}
  \label{fig:RWREAssembly}
\end{figure}

\medskip\noindent\emph{Comparison of Numerical and Theoretical Results---}
Fig. \ref{fig:MaxVar} and \ref{fig:QuantileVar} show that the asymptotic theoretical predictions for $\var{\maxnt}$ and $\var{\envnt}$ are in excellent agreement with the numerical measurements.  Fig. \ref{fig:QuantileVar} further shows that we reliably recover $\var{\envnt}$ using $\varnum{\maxnt} - \varasy{\snt}$, as expected from Eq. \ref{eq:varadd}. Notably, while these results were derived for asymptotically large $\log(N)$ and $t$, they hold nearly perfectly down to $N=10^{2}$.
Fig. \ref{fig:QuantileVar} reveals that while we readily see the long-time $t^{1/2}$ power-law for $\var{\envnt}$ from Eq. \ref{eq:varqnt}, the $t^{1/3}$ power-law is elusive. Although the full characterization of the short-time regime is in excellent agreement with the numerical results, the $t^{1/3}$ power-law is difficult to capture since the transitional window of $\log(N)$ to $(\log(N))^2$ is too narrow for realistic sizes of $N$, even up to $N=10^{300}$. By measuring the long-time $t^{1/2}$ power law, we measure the short-time scaling behavior of the KPZ equation up to a prefactor using Eq. \ref{eqlogNsq}. Fig. \ref{fig:RWREAssembly} shows the tight matching of the asymptotic theory curves and numerically measured values for the variance of $\maxnt, \envnt$ and $\snt$ for a given value of $N=10^7$. Notice that for $t\approx \log(N)$ the asymptotic theory and numerical values for the variance of $\snt$ do not fit as well as for large $t$. This is likely a result of finite-size effects and quickly goes away at larger values of $t$ or when $N$ increases. The fit for $N=10^{300}$ in Fig. \ref{fig:MaxVar} and \ref{fig:QuantileVar} remains tight over the entire range of $t$.

\medskip\noindent\emph{Conclusion---}
The link between RWREs and KPZ universality with its wealth of theoretical, numerical and experimental evidence strongly suggests that aspects of the picture presented here will persist beyond discrete and solvable models, even to experiments. When $t$ is of order $\log(N)$, variances should be non-universal, depending in a difficult to determine way on the nature of the environment. By contrast, when $t\gg\log(N)$, we anticipate that the scaling exponents and functional forms we have identified for the variances of $\envnt,\snt$ and $\maxnt$ will be universal, as will the relation \eqref{eq:varadd}. The leading coefficients  in Eq. \eqref{eq:varqnt} should be non-universal and hold within them all of the accessible information about the correlation structure of the environment -- we call these {\it extreme diffusion coefficients}. Further theoretical study, such as for the general $\alpha,\beta$ Beta RWRE model, should provide a natural first test of this universal picture and an understanding of how the extreme diffusion coefficients relate to the microscopic environment. A continuum model that should provide an even wider testing-ground amenable to numerics involves particles $x_i(t)$ for $i=1,2,\ldots$ satisfying $dx_i(t) = F(x_i(t),t)dt + D(x_i(t),t) dB_i(t)$ where $F(x,t)$ and $D(x,t)$ are random forcing (as in \cite{ledoussalDiffusionTimedependentRandom2017}) and diffusivity (generalizing diffusing diffusivity, c.f. \cite{PhysRevX.7.021002}) fields common to all particles while $B_i$ are  Brownian motions independent between different $i$. Changing the correlation structures of $F$ and $D$ will probe the transition between temporally mixing versus quenched environments, which should have very different behavior (c.f. \cite{PhysRevLett.61.500,PhysRevLett.62.3097}) and warrants further study. Considering higher dimensions as in~\cite{ledoussalDiffusionTimedependentRandom2017} may lead to further theory that better model real physical systems. A study of higher order cumulants may reveal other ways to probe the hidden environment, although they may be harder to observe numerically or experimentally.

In physical systems it is impossible to directly measure the environmental variance. However, an indirect measurement can be performed via the approach presented here by using $\var \envnt\approx \var \maxnt - \var \snt$. The sample variance $\var \snt$ is now computed using
$
 \var \snt = \frac{\pi^2 D}{6} \frac{t}{\log(N)}
$
where $D$ is the diffusion coefficient. One could repeatedly track the motion of the leading edge of diffusing particles in a system of colloids confined to a quasi-1D channel thereby directly measuring $\var \maxnt$ for system sizes ranging from $N ~ 10^2$ to $N ~ 10^{10}$. Further, one can also perform complementary measurements on the time of first passage of diffusing objects, which opens the door to experiments done on all manner of diffusing objects, including light or sound diffusing through a scattering medium, dye molecules in a fluid, or any other object whose first passage can be measured.
By measuring the environmental variance and extreme diffusion coefficient we will gain a new microscope through which to probe the hidden nature of the underlying environment in which the diffusion occurs. Our work should serve as a guide in the development and analysis of novel experimental measurements of the extreme behavior of many-particle diffusion.

\medskip\noindent\emph{Acknowledgements---}
We thank G. Barraquand and P. Le Doussal for discussions and S. Prolhac for providing numerics for $\var{h(0,s)}$.
This work was funded under the W.M. Keck Foundation Science and Engineering grant on ``Extreme Diffusion''.
I.C. also wishes to acknowledge ongoing support from the NSF through DMS:1811143 and DMS:1937254, the Simon Foundation through a Simons Fellowship in Mathematics (Grant No. 817655), and the Packard Foundation Fellowship for Science and Engineering. Much of this work was performed while I.C. held a Miller Visiting Professorship from the Miller Institute for Basic Research in Science, and while in residence at the Mathematical Sciences Research Institute in Berkeley, California (NSF Grant No. 1440140).
E.I.C. wishes to acknowledge ongoing support from the Simons Foundation for the collaboration Cracking the Glass Problem via award 454939. This work benefited from access to the University of Oregon high performance computing cluster, Talapas.

\bibliographystyle{unsrt}
\bibliography{main}

\begin{thebibliography}{10}

\bibitem{saxtonSingleparticleTrackingApplications1997a}
Michael~J. Saxton and Ken Jacobson.
\newblock Single-particle tracking: {{Applications}} to membrane dynamics.
\newblock {\em Annual Review of Biophysics and Biomolecular Structure},
  26:373--399, 1997.

\bibitem{pintoPhysicsTypeIa2000}
Philip~A. Pinto and Ronald~G. Eastman.
\newblock The {{Physics}} of {{Type Ia Supernova Light Curves}}. {{II}}.
  {{Opacity}} and {{Diffusion}}.
\newblock {\em The Astrophysical Journal}, 530(2):757--776, February 2000.

\bibitem{hoflingAnomalousTransportCrowded2013a}
Felix H{\"o}fling and Thomas Franosch.
\newblock Anomalous transport in the crowded world of biological cells.
\newblock {\em Reports on Progress in Physics}, 76(4), 2013.

\bibitem{ghoshNonuniversalTracerDiffusion2015}
Surya~K. Ghosh, Andrey~G. Cherstvy, and Ralf Metzler.
\newblock Non-universal tracer diffusion in crowded media of non-inert
  obstacles.
\newblock {\em Physical Chemistry Chemical Physics}, 17(3):1847--1858, January
  2015.

\bibitem{manzoReviewProgressSingle2015}
Carlo Manzo and Maria~F. {Garcia-Parajo}.
\newblock A review of progress in single particle tracking: {{From}} methods to
  biophysical insights.
\newblock {\em Reports on Progress in Physics}, 78(12), 2015.

\bibitem{iyer-biswasFirstPassageProcesses2016}
Srividya {Iyer-Biswas} and Anton Zilman.
\newblock First {{Passage}} processes in cellular biology.
\newblock {\em Advances in Chemical Physics}, 160:261--306, April 2016.

\bibitem{metzlerNonBrownianDiffusionLipid2016c}
R.~Metzler, J.-H. Jeon, and A.~G. Cherstvy.
\newblock Non-{{Brownian}} diffusion in lipid membranes: {{Experiments}} and
  simulations.
\newblock {\em Biochimica Et Biophysica Acta}, 1858(10):2451--2467, October
  2016.

\bibitem{shenSingleParticleTracking2017}
Hao Shen, Lawrence~J. Tauzin, Rashad Baiyasi, Wenxiao Wang, Nicholas Moringo,
  Bo~Shuang, and Christy~F. Landes.
\newblock Single {{Particle Tracking}}: {{From Theory}} to {{Biophysical
  Applications}}.
\newblock {\em Chemical Reviews}, 117(11):7331--7376, June 2017.

\bibitem{hufnagelForecastControlEpidemics2004a}
L.~Hufnagel, D.~Brockmann, and T.~Geisel.
\newblock Forecast and control of epidemics in a globalized world.
\newblock {\em Proceedings of the National Academy of Sciences},
  101(42):15124--15129, October 2004.

\bibitem{kaoVirusDiffusionIsolation2006b}
P.~H. Kao and R.~J. Yang.
\newblock Virus diffusion in isolation rooms.
\newblock {\em Journal of Hospital Infection}, 62(3):338--345, March 2006.

\bibitem{ewensMathematicalPopulationGenetics2012}
Warren~John Ewens.
\newblock {\em Mathematical {{Population Genetics}}, {{I}}. {{Theoretical}}
  Introduction}, volume~27.
\newblock 2012.

\bibitem{metzlerFirstpassagePhenomenaTheir2014}
Ralf Metzler, Gleb Oshanin, and Sidney Redner.
\newblock {\em First-Passage Phenomena and Their Applications}.
\newblock {World Scientific Publishing Co.}, January 2014.

\bibitem{waxmanDiffusionEquationRandom2017}
David Waxman.
\newblock The diffusion equation of random genetic drift\textendash biology's
  analogue of the {{Schr\"odinger}} equation?
\newblock {\em Contemporary Physics}, 58(3):253--261, July 2017.

\bibitem{einsteinUberMolekularkinetischenTheorie1905b}
A.~Einstein.
\newblock \"uber die von der molekularkinetischen {{Theorie}} der {{W\"arme}}
  geforderte {{Bewegung}} von in ruhenden {{Fl\"ussigkeiten}} suspendierten
  {{Teilchen}}.
\newblock {\em Annalen der Physik}, 322(8):549--560, 1905.

\bibitem{einsteinZurTheorieBrownschen1906a}
A.~Einstein.
\newblock Zur {{Theorie}} der {{Brownschen Bewegung}}.
\newblock {\em Annalen der Physik}, 324(2):371--381, 1906.

\bibitem{einsteinTheoretischeBemerkungenUber1907a}
A.~Einstein.
\newblock Theoretische {{Bemerkungen \"Uber}} die {{Brownsche Bewegung}}.
\newblock {\em Zeitschrift f\"ur Elektrochemie und angewandte physikalische
  Chemie}, 13(6):41--42, 1907.

\bibitem{rassoul-aghaQuenchedFreeEnergy2013}
Firas {Rassoul-Agha}, Timo Sepp{\"a}l{\"a}inen, and Atilla Yilmaz.
\newblock Quenched {{Free Energy}} and {{Large Deviations}} for {{Random
  Walks}} in {{Random Potentials}}.
\newblock {\em Communications on Pure and Applied Mathematics}, 66(2):202--244,
  2013.

\bibitem{brownXXVIIBriefAccount1828a}
Robert Brown.
\newblock {{XXVII}}. {{A}} brief account of microscopical observations made in
  the months of {{June}}, {{July}} and {{August}} 1827, on the particles
  contained in the pollen of plants; and on the general existence of active
  molecules in organic and inorganic bodies.
\newblock {\em The Philosophical Magazine}, 4(21):161--173, September 1828.

\bibitem{brownXXIVAdditionalRemarks1829a}
Robert Brown.
\newblock {{XXIV}}. {{Additional}} remarks on active molecules.
\newblock {\em The Philosophical Magazine}, 6(33):161--166, September 1829.

\bibitem{langevinTheorieMouvementBrownien1908a}
Paul Langevin.
\newblock Sur la theorie du mouvement brownien.
\newblock {\em Compt. Rendus}, 146:530--533, 1908.

\bibitem{sutherlandLIIViscosityGases1893}
William Sutherland.
\newblock {{LII}}. {{The}} viscosity of gases and molecular force.
\newblock {\em The London, Edinburgh, and Dublin Philosophical Magazine and
  Journal of Science}, 36(223):507--531, December 1893.

\bibitem{sutherlandLXXVDynamicalTheory1905}
William Sutherland.
\newblock {{LXXV}}. {{A}} dynamical theory of diffusion for non-electrolytes
  and the molecular mass of albumin.
\newblock {\em The London, Edinburgh, and Dublin Philosophical Magazine and
  Journal of Science}, 9(54):781--785, June 1905.

\bibitem{vonsmoluchowskiZurKinetischenTheorie1906}
M.~{von Smoluchowski}.
\newblock Zur kinetischen {{Theorie}} der {{Brownschen Molekularbewegung}} und
  der {{Suspensionen}}.
\newblock {\em Annalen der Physik}, 326(14):756--780, 1906.

\bibitem{vonsmoluchowskiNotizUiberBerechnung1915}
M.~Von~Smoluchowski.
\newblock Notiz uiber die {{Berechnung}} der {{Brownschen Molekularbewegung}}
  bei der {{Ehrenhaft-Millikanschen Versuchsanordning}}.
\newblock {\em Phys. Z}, 16:318--321, 1915.

\bibitem{perrinMouvementBrownienRealite1909a}
Jean~Baptiste Perrin.
\newblock Le {{Mouvement Brownien}} et la {{R\'ealit\'e Moleculaire}}.
\newblock {\em Ann. Chimi. Phys.}, 18(8):5--114, 1909.

\bibitem{perrinMouvementBrownienMolecules1910a}
Jean Perrin.
\newblock {Mouvement brownien et mol\'ecules}.
\newblock {\em Journal de Chimie Physique}, 8:57--91, 1910.

\bibitem{uhlenbeckTheoryBrownianMotion1930a}
G.~E. Uhlenbeck and L.~S. Ornstein.
\newblock On the {{Theory}} of the {{Brownian Motion}}.
\newblock {\em Physical Review}, 36(5):823--841, September 1930.

\bibitem{huangDirectObservationFull2011}
Rongxin Huang, Isaac Chavez, Katja~M. Taute, Branimir Luki{\'c}, Sylvia Jeney,
  Mark~G. Raizen, and Ernst-Ludwig Florin.
\newblock Direct observation of the full transition from ballistic to diffusive
  {{Brownian}} motion in a liquid.
\newblock {\em Nature Physics}, 7(7):576--580, July 2011.

\bibitem{hammondDirectMeasurementBallistic2017}
Andrew~P. Hammond and Eric~I. Corwin.
\newblock Direct measurement of the ballistic motion of a freely floating
  colloid in {{Newtonian}} and viscoelastic fluids.
\newblock {\em Physical Review E}, 96(4):042606--042606, October 2017.

\bibitem{lukicDirectObservationNondiffusive2005}
B.~Luki{\'c}, S.~Jeney, C.~Tischer, A.~J. Kulik, L.~Forr{\'o}, and E.-L.
  Florin.
\newblock Direct {{Observation}} of {{Nondiffusive Motion}} of a {{Brownian
  Particle}}.
\newblock {\em Physical Review Letters}, 95(16):160601, October 2005.

\bibitem{franoschResonancesArisingHydrodynamic2011}
Thomas Franosch, Matthias Grimm, Maxim Belushkin, Flavio~M. Mor, Giuseppe
  Foffi, L{\'a}szl{\'o} Forr{\'o}, and Sylvia Jeney.
\newblock Resonances arising from hydrodynamic memory in {{Brownian}} motion.
\newblock {\em Nature}, 478(7367):85--88, October 2011.

\bibitem{kheifetsObservationBrownianMotion2014}
Simon Kheifets, Akarsh Simha, Kevin Melin, Tongcang Li, and Mark~G. Raizen.
\newblock Observation of {{Brownian Motion}} in {{Liquids}} at {{Short Times}}:
  {{Instantaneous Velocity}} and {{Memory Loss}}.
\newblock {\em Science}, 343(6178):1493--1496, March 2014.

\bibitem{wangWhenBrownianDiffusion2012a}
Bo~Wang, James Kuo, Sung~Chul Bae, and Steve Granick.
\newblock When {{Brownian}} diffusion is not {{Gaussian}}.
\newblock {\em Nature Materials}, 11(6):481--485, May 2012.

\bibitem{bouchaudAnomalousDiffusionDisordered1990a}
Jean~Philippe Bouchaud and Antoine Georges.
\newblock Anomalous diffusion in disordered media: {{Statistical}} mechanisms,
  models and physical applications.
\newblock {\em Physics Reports}, 195(4-5):127--293, November 1990.

\bibitem{metzlerBrownianMotionFirstpassage2019a}
Ralf Metzler.
\newblock Brownian motion and beyond: First-passage, power spectrum,
  non-{{Gaussianity}}, and anomalous diffusion.
\newblock {\em Journal of Statistical Mechanics: Theory and Experiment},
  2019(11):114003--114003, November 2019.

\bibitem{ramaswamyMechanicsStatisticsActive2010b}
Sriram Ramaswamy.
\newblock The {{Mechanics}} and {{Statistics}} of {{Active Matter}}.
\newblock {\em Annual Review of Condensed Matter Physics}, 1(1):323--345, 2010.

\bibitem{kanazawaLoopyLevyFlights2020a}
Kiyoshi Kanazawa, Tomohiko~G. Sano, Andrea Cairoli, and Adrian Baule.
\newblock Loopy {{L\'evy}} flights enhance tracer diffusion in active
  suspensions.
\newblock {\em Nature}, 579(7799):364--367, March 2020.

\bibitem{zangiFrequencydependentStokesEinsteinRelation2007}
Ronen Zangi and Laura~J. Kaufman.
\newblock Frequency-dependent {{Stokes-Einstein}} relation in supercooled
  liquids.
\newblock {\em Physical Review E}, 75(5):051501, May 2007.

\bibitem{chernovReplicationMulticomponentChain1967}
A.~A. Chernov.
\newblock Replication of a multicomponent chain by the "lightning" mechanism.
\newblock {\em Biophysics}, 12(2), 1967.

\bibitem{temkinOnedimensionalRandomWalks1972}
D.~E. Temkin.
\newblock One-dimensional random walks in a two-component chain.
\newblock {\em Soviet Math. Docl.}, 13:1172--1176, 1972.

\bibitem{havlinDiffusionDisorderedMedia1987}
Shlomo Havlin and Daniel {Ben-Avraham}.
\newblock Diffusion in disordered media.
\newblock {\em Advances in Physics}, 36(6):695--798, January 1987.

\bibitem{bolthausenTenLecturesRandom2002}
Erwin Bolthausen and Alain-Sol Sznitman.
\newblock {\em Ten Lectures on Random Media}.
\newblock {Birkh\"auser Basel}, 2002.

\bibitem{sznitmanTopicsRandomWalks2004}
A.-S. Sznitman.
\newblock Topics in random walks in random environment.
\newblock Technical Report 92-95003-25-X, {International Atomic Energy Agency
  (IAEA)}, 2004.

\bibitem{oferRandomWalksRandom2004}
Zeitouni Ofer.
\newblock Random walks in random environment.
\newblock In {\em Lectures on Probability Thorey and Statistics, {{Lecture
  Notes}}}, volume 1837 of {\em Math}, pages 189--312. {Springer}, {Berlin},
  2004.

\bibitem{hughesRandomWalksRandom1995}
Barry~D. Hughes.
\newblock {\em Random {{Walks}} and {{Random Environments}}: {{Random}} Walks}.
\newblock {Clarendon Press}, 1995.

\bibitem{kestenLimitLawRandom1975}
H.~Kesten, M.~V. Kozlov, and F.~Spitzer.
\newblock {A limit law for random walk in a random environment}.
\newblock {\em Compositio Mathematica}, 30(2):145--168, 1975.

\bibitem{sinaiLimitingBehaviorOneDimensional1983a}
Ya.~G. Sinai.
\newblock The {{Limiting Behavior}} of a {{One-Dimensional Random Walk}} in a
  {{Random Medium}}.
\newblock {\em Theory of Probability \& Its Applications}, 27(2):256--268,
  January 1983.

\bibitem{bouchaudClassicalDiffusionParticle1990a}
J.~P Bouchaud, A~Comtet, A~Georges, and P~Le~Doussal.
\newblock Classical diffusion of a particle in a one-dimensional random force
  field.
\newblock {\em Annals of Physics}, 201(2):285--341, August 1990.

\bibitem{burlatskyTransientRelaxationCharged1998}
S.~F. Burlatsky and John~M. Deutch.
\newblock Transient relaxation of a charged polymer chain subject to an
  external field in a random tube.
\newblock {\em The Journal of Chemical Physics}, 109(6):2572--2578, August
  1998.

\bibitem{ledoussalRandomWalkersOnedimensional1999}
Pierre Le~Doussal, C{\'e}cile Monthus, and Daniel~S. Fisher.
\newblock Random walkers in one-dimensional random environments: {{Exact}}
  renormalization group analysis.
\newblock {\em Physical Review E}, 59(5):4795--4840, May 1999.

\bibitem{richardsonAtmosphericDiffusionShown1926}
Lewis~Fry Richardson and Gilbert~Thomas Walker.
\newblock Atmospheric diffusion shown on a distance-neighbour graph.
\newblock {\em Proceedings of the Royal Society of London. Series A, Containing
  Papers of a Mathematical and Physical Character}, 110(756):709--737, April
  1926.

\bibitem{hentschelRelativeDiffusionTurbulent1984}
H.~G.~E. Hentschel and Itamar Procaccia.
\newblock Relative diffusion in turbulent media: {{The}} fractal dimension of
  clouds.
\newblock {\em Physical Review A}, 29(3):1461--1470, March 1984.

\bibitem{bouchaudDiffusionLocalizationWaves1990}
J.~P. Bouchaud.
\newblock Diffusion and {{Localization}} of {{Waves}} in a {{Time-Varying
  Random Environment}}.
\newblock {\em Europhysics Letters (EPL)}, 11(6):505--510, March 1990.

\bibitem{chertkovAnomalousScalingExponents1996}
M.~Chertkov and G.~Falkovich.
\newblock Anomalous {{Scaling Exponents}} of a {{White-Advected Passive
  Scalar}}.
\newblock {\em Physical Review Letters}, 76(15):2706--2709, April 1996.

\bibitem{bernardAnomalousScalingNPoint1996}
Denis Bernard, Krzysztof Gawedzki, and Antti Kupiainen.
\newblock Anomalous {{Scaling}} in the {{N-Point Functions}} of {{Passive
  Scalar}}.
\newblock {\em Physical Review E}, 54(3):2564--2572, September 1996.

\bibitem{jullienRichardsonPairDispersion1999a}
Marie-Caroline Jullien, J{\'e}r{\^o}me Paret, and Patrick Tabeling.
\newblock Richardson {{Pair Dispersion}} in {{Two-Dimensional Turbulence}}.
\newblock {\em Physical Review Letters}, 82(14):2872--2875, April 1999.

\bibitem{balkovskyIntermittentDistributionInertial2001}
E.~Balkovsky, G.~Falkovich, and A.~Fouxon.
\newblock Intermittent {{Distribution}} of {{Inertial Particles}} in
  {{Turbulent Flows}}.
\newblock {\em Physical Review Letters}, 86(13):2790--2793, March 2001.

\bibitem{rassoul-aghaAlmostSureInvariance2005}
F.~{Rassoul-Agha} and T.~Seppalainen.
\newblock An almost sure invariance principle for random walks in a space-time
  random environment.
\newblock {\em Probability Theory and Related Fields}, 133(3):299--314,
  November 2005.

\bibitem{deuschelQuenchedInvariancePrinciple2016}
Jean-Dominique Deuschel, Xiaoqin Guo, and Alejandro~F. Ramirez.
\newblock Quenched invariance principle for random walk in time-dependent
  balanced random environment.
\newblock {\em arXiv:1503.01964 [math]}, September 2016.

\bibitem{balazsRandomAverageProcess2006}
Marton Balazs, Firas {Rassoul-Agha}, and Timo Seppalainen.
\newblock The random average process and random walk in a space-time random
  environment in one dimension.
\newblock {\em Communications in Mathematical Physics}, 266(2):499--545,
  September 2006.

\bibitem{barraquandRandomwalkBetadistributedRandom2017a}
Guillaume Barraquand and Ivan Corwin.
\newblock Random-walk in {{Beta-distributed}} random environment.
\newblock {\em Probability Theory and Related Fields}, 167(3-4):1057--1116,
  April 2017.

\bibitem{corwinKardarParisiZhang2012a}
Ivan Corwin.
\newblock The kardar\textendash parisi\textendash zhang equation and
  universality class.
\newblock {\em Random Matrices: Theory and Applications}, 01(01):1130001,
  January 2012.

\bibitem{quastelOneDimensionalKPZEquation2015a}
Jeremy Quastel and Herbert Spohn.
\newblock The {{One-Dimensional KPZ Equation}} and {{Its Universality Class}}.
\newblock {\em Journal of Statistical Physics}, 160(4):965--984, August 2015.

\bibitem{tracyLevelspacingDistributionsAiry1993}
Craig~A. Tracy and Harold Widom.
\newblock Level-spacing distributions and the {{Airy}} kernel.
\newblock {\em Physics Letters B}, 305(1):115--118, May 1993.

\bibitem{ledoussalDiffusionTimedependentRandom2017}
Pierre Le~Doussal and Thimoth{\'e}e Thiery.
\newblock Diffusion in time-dependent random media and the
  {{Kardar-Parisi-Zhang}} equation.
\newblock {\em Physical Review E}, 96(1):010102(R), July 2017.

\bibitem{barraquandModerateDeviationsDiffusion2020a}
Guillaume Barraquand and Pierre~Le Doussal.
\newblock Moderate deviations for diffusion in time dependent random media.
\newblock {\em Journal of Physics A: Mathematical and Theoretical},
  53(21):215002, May 2020.

\bibitem{kardarDynamicScalingGrowing1986a}
Mehran Kardar, Giorgio Parisi, and Yi-Cheng Zhang.
\newblock Dynamic {{Scaling}} of {{Growing Interfaces}}.
\newblock {\em Physical Review Letters}, 56(9):889--892, March 1986.

\bibitem{sasamotoOneDimensionalKardarParisiZhangEquation2010}
Tomohiro Sasamoto and Herbert Spohn.
\newblock One-{{Dimensional Kardar-Parisi-Zhang Equation}}: {{An Exact
  Solution}} and its {{Universality}}.
\newblock {\em Physical Review Letters}, 104(23):230602, June 2010.

\bibitem{calabreseFreeenergyDistributionDirected2010}
P.~Calabrese, P.~Le Doussal, and A.~Rosso.
\newblock Free-energy distribution of the directed polymer at high temperature.
\newblock {\em EPL (Europhysics Letters)}, 90(2):20002, April 2010.

\bibitem{dotsenkoBetheAnsatzDerivation2010}
V.~Dotsenko.
\newblock Bethe ansatz derivation of the {{Tracy-Widom}} distribution for
  one-dimensional directed polymers.
\newblock {\em EPL (Europhysics Letters)}, 90(2):20003, April 2010.

\bibitem{amirProbabilityDistributionFree2011}
Gideon Amir, Ivan Corwin, and Jeremy Quastel.
\newblock Probability {{Distribution}} of the {{Free Energy}} of the
  {{Continuum Directed Random Polymer}} in 1+1 dimensions.
\newblock {\em Communications on Pure and Applied Mathematics}, 64(4):466--537,
  April 2011.

\bibitem{sabotRandomWalksDirichlet2017}
Christophe Sabot and Laurent Tournier.
\newblock Random walks in {{Dirichlet}} environment: An overview.
\newblock {\em Annales de la Facult\'e des sciences de Toulouse :
  Math\'ematiques}, 26(2):463--509, 2017.

\bibitem{balazsLargeDeviationsWandering2019}
M{\'a}rton Bal{\'a}zs, Firas {Rassoul-Agha}, and Timo Sepp{\"a}l{\"a}inen.
\newblock Large deviations and wandering exponent for random walk in a dynamic
  beta environment.
\newblock {\em The Annals of Probability}, 47(4):2186--2229, July 2019.

\bibitem{barraquandLargeDeviationsSticky2020}
Guillaume Barraquand and Mark Rychnovsky.
\newblock Large deviations for sticky {{Brownian}} motions.
\newblock {\em Electronic Journal of Probability}, 25(none):1--52, January
  2020.

\bibitem{brockingtonBetheAnsatzSticky2021}
Dom Brockington and Jon Warren.
\newblock The {{Bethe Ansatz}} for {{Sticky Brownian Motions}}.
\newblock {\em arXiv:2104.06482 [math]}, April 2021.

\bibitem{oviedoSecondOrderFluctuations2021}
Giancarlos Oviedo, Gonzalo Panizo, and Alejandro~F. Ram{\'i}rez.
\newblock Second order fluctuations of large deviations for perturbed random
  walks.
\newblock {\em arXiv:2108.02877 [math]}, August 2021.

\bibitem{korotkikhHiddenDiagonalIntegrability2022}
Sergei Korotkikh.
\newblock Hidden diagonal integrability of q-{{Hahn}} vertex model and {{Beta}}
  polymer model.
\newblock {\em Probability Theory and Related Fields}, March 2022.

\bibitem{krajenbrinkCrossoverMacroscopicFluctuation2022}
Alexandre Krajenbrink and Pierre~Le Doussal.
\newblock The crossover from the {{Macroscopic Fluctuation Theory}} to the
  {{Kardar-Parisi-Zhang}} equation controls the large deviations beyond
  {{Einstein}}'s diffusion.
\newblock {\em arXiv:2204.04720 [cond-mat, physics:math-ph, physics:nlin]},
  April 2022.

\bibitem{albertsIntermediateDisorderRegime2010}
Tom Alberts, Kostya Khanin, and Jeremy Quastel.
\newblock Intermediate {{Disorder Regime}} for {{Directed Polymers}} in
  {{Dimension}} 1+1.
\newblock {\em Physical Review Letters}, 105(9):090603, August 2010.

\bibitem{PhysRevA.45.638}
Joachim Krug, Paul Meakin, and Timothy Halpin-Healy.
\newblock Amplitude universality for driven interfaces and directed polymers in
  random media.
\newblock {\em Phys. Rev. A}, 45:638--653, Jan 1992.

\bibitem{prolhacHeightDistributionKPZ2011}
Sylvain Prolhac and Herbert Spohn.
\newblock The height distribution of the {{KPZ}} equation with sharp wedge
  initial condition: Numerical evaluations.
\newblock {\em Physical Review E}, 84(1):011119, July 2011.

\bibitem{halpin-healyKPZCocktailShakenNot2015a}
Timothy {Halpin-Healy} and Kazumasa~A. Takeuchi.
\newblock A {{KPZ Cocktail-Shaken}}, not {{Stirred}}...
\newblock {\em Journal of Statistical Physics}, 160(4):794--814, August 2015.

\bibitem{fellerDiffusionProcessesGenetics1951}
William Feller.
\newblock Diffusion {{Processes}} in {{Genetics}}.
\newblock {\em Proceedings of the Second Berkeley Symposium on Mathematical
  Statistics and Probability}, 2:227--247, January 1951.

\bibitem{moranRandomProcessesGenetics1958}
P.~a.~P. Moran.
\newblock Random processes in genetics.
\newblock {\em Mathematical Proceedings of the Cambridge Philosophical
  Society}, 54(1):60--71, January 1958.

\bibitem{pollardGaseousSelfDiffusionLong1948}
W.~G. Pollard and R.~D. Present.
\newblock On {{Gaseous Self-Diffusion}} in {{Long Capillary Tubes}}.
\newblock {\em Physical Review}, 73(7):762--774, April 1948.

\bibitem{ahmadiDiffusionQuasionedimensionalChannels2017}
Sheida Ahmadi and Richard~K. Bowles.
\newblock Diffusion in quasi-one-dimensional channels: {{A}} small system n, p,
  {{T}}, transition state theory for hopping times.
\newblock {\em The Journal of Chemical Physics}, 146(15):154505, April 2017.

\bibitem{SeeSupplementalMaterial}
See {{Supplemental Material}} at {{XXXX}}.

\bibitem{prahoferUniversalDistributionsGrowth2000}
Michael Pr{\"a}hofer and Herbert Spohn.
\newblock Universal {{Distributions}} for {{Growth Processes}} in 1+1
  {{Dimensions}} and {{Random Matrices}}.
\newblock {\em Physical Review Letters}, 84(21):4882--4885, May 2000.

\bibitem{PhysRevX.7.021002}
Aleksei~V. Chechkin, Flavio Seno, Ralf Metzler, and Igor~M. Sokolov.
\newblock Brownian yet non-gaussian diffusion: From superstatistics to
  subordination of diffusing diffusivities.
\newblock {\em Phys. Rev. X}, 7:021002, Apr 2017.

\bibitem{PhysRevLett.61.500}
S.~H. Noskowicz and I.~Goldhirsch.
\newblock Average versus typical mean first-passage time in a random random
  walk.
\newblock {\em Phys. Rev. Lett.}, 61:500--502, Aug 1988.

\bibitem{PhysRevLett.62.3097}
Pierre Le~Doussal.
\newblock First-passage time for random walks in random environments.
\newblock {\em Phys. Rev. Lett.}, 62:3097--3097, Jun 1989.

\end{thebibliography}

\end{document}